\documentclass[reprint,amsmath,amssymb,aps,nofootinbib]{revtex4-1}

\usepackage{graphicx}
\usepackage{dcolumn}
\usepackage{bm}
\usepackage[normalem]{ulem}

\usepackage{color}
\usepackage{xcolor}

\usepackage{mathrsfs}
\usepackage{amsfonts}
\usepackage{varioref}
\usepackage{csquotes}
\usepackage{hyphenat}
\usepackage{float}
\usepackage{multirow}

\RequirePackage[colorlinks,citecolor=blue,urlcolor=magenta,linkcolor=blue]{hyperref}

\def \S {Schwarzschild}

\def\rstar{r_{\star}}
\DeclareMathOperator{\sech}{sech}
\def\rh{r_{_{H}}}
\def\rs{r_{s}}
\def\ms{m_{\star}}
\labelformat{chapter}{Chapter (#1)} 
\labelformat{section}{Section (#1)} 
\labelformat{subsection}{Section (#1)} 
\labelformat{subsubsection}{Section (#1)}
\labelformat{subsubsubsection}{Section (#1)}
\labelformat{equation}{Eq.~(#1)} 
\labelformat{figure}{Fig.~(#1)} 
\labelformat{subfigure}{Fig.~(\thefigure#1)} 
\labelformat{table}{Tab.~(#1)} 
\labelformat{appendix}{Appendix #1}
\begin{document}

\title{Thermal behavior of a radially deformed black hole spacetime}

\author{Subhajit Barman} 
\email{subhajit.b@iitg.ac.in}
\affiliation{Department of Physics, Indian Institute of Technology Guwahati, Guwahati 781039, Assam, India}

\author{Sajal Mukherjee}
\email{sajal@iucaa.in}
\affiliation{Inter-University Centre for Astronomy and Astrophysics, Post Bag 4, Pune-411007, India}

\begin{abstract}
\noindent
In the present article, we study the Hawking effect and the bounds on greybody factor in a spacetime with radial deformation. This deformation is expected to carry the imprint of a non-Einsteinian theory of gravity, but shares some of the important characteristics of general relativity (GR). In particular, this radial deformation will restore the asymptotic behavior, and also allows for the separation of the scalar field equation in terms of the angular and radial coordinates --- making it suitable to study the Hawking effect and greybody factors. However, the radial deformation would introduce a change in the locations of the horizon, and therefore, the temperature of the Hawking effect naturally alters. In fact, we observe that the deformation parameter has an enhancing effect on both temperature and bounds on the greybody factor, which introduces a useful distinction with the Kerr spacetime. We discuss these effects elaborately, and broadly study the thermal behavior of a radially deformed spacetime.
\end{abstract}
\pacs{}
\maketitle
\section{Introduction}
The Kerr metric is one of the remarkable findings of GR \cite{Kerr:1963ud}. From the weak field to strong field regime, Kerr solution has passed all tests with flying colors \cite{Hulse:1974eb,Smoot:1992td,Perlmutter:1998np,Everitt:2011hp}. All of these success stories make a strong case of GR, and even constrained some of the alternative theories of gravity \cite{Mukherjee:2017fqz,berti2015testing}. Moreover, with the gravitational wave (GW) astronomy coming to the fore, these studies emerge with brighter possibilities \cite{Abbott:2016blz,TheLIGOScientific:2017qsa,LIGOScientific:2018mvr,Abbott:2020khf}. Besides these successes of GR, there are also strong limitations which motivates to seek for alternatives. It is also known that GR fails to explain both the small and large scale structure of the \emph{nature}, and therefore, a modification of the theory is required \cite{Riess:1998cb,Hawking:1976ra}. 

If we attempt to modify GR, it is also bound to happen that the resultant spacetime may loose some of its useful properties, such as axis-symmetry or separability condition. Nonetheless, if we assure that these conditions are built-in restored, and still aim to modify spacetime structure, we may end up with constraining its metric functions \cite{Konoplya:2018arm}. The deviation from GR may be coded within these constraints. One of such possibilities comes into play if we modify $\Delta$, which in Kerr case is $\Delta=r^2- r \rs+a^2$, and $\rs=2M$ with $M$ being the mass of the black hole (BH). In the present article, we will be concerned with this specific example, where we modify $\Delta$ by adding a $r$-dependent term to it \cite{Konoplya:2016pmh}. Note that this deformation is only radial, and does not effect angular distribution of the spacetime. 

The motivation to study a spacetime which mimics a  radially deformed Kerr spacetime is two folded. First, it provides a simple yet useful extension of GR, which is well-grounded with GW data \cite{ Abbott:2016nmj}. Therefore, it can be a potential candidate of alternative theories of gravity, and studying along this line can be beneficial. The metric corresponding to these deformed BH spacetimes has the same asymptotic features as the original ones from Einstein gravity. Secondly, the radial deformation makes it possible for the field equation to be separable in terms of the radial and angular coordinates, which paves the way for the formulation of semi-classical analysis in these spacetimes. Furthermore, the horizon structure differs, as the position of the horizon is now changed due to the introduction of the deformation. Then the effects of these deformations will also be felt through the predictions of semi-classical gravity, which concerns the horizon structure. In this regard, the Hawking effect \cite{hawking1975} is a major arena to venture in, which states that an asymptotic observer in a BH spacetime will realize a Planckian thermal distribution of particles with temperature proportional to the surface gravity of the BH's event horizon. In the deformed BH spacetime, the distortion in the horizon structure is expected to change its surface gravity, which naturally affects the spectrum of the perceived Hawking radiation. 

Another important thing to note, is that the spectrum of the Hawking effect as should practically be seen by an asymptotic observer is not an absolute blackbody distribution, rather it is a greybody distribution. This greybody distribution is characterized by the transmission coefficient through the effective potential of the considered field. Greybody factor also contains the information regarding different BH parameters. Here also one can expect prominent effects of the  deformation parameter. However, a straight forward exact estimation of these greybody factors is an insuperable job analytically, though one can seek the help of numerical methods \cite{Harris:2003eg, Rocha:2009xy, Catalan:2014ama, Becar:2014aka, Dong:2015qpa, Pappas:2016ovo, Gray:2015pma, Abedi:2013xua}. Analytically these estimations can be performed in asymptotic frequency regimes \cite{Harmark:2007jy, Keshet:2007be, Kim:2007gj, Rocha:2009xy, Gonzalez:2010vv, Gonzalez:2010ht, Kanti:2014dxa, CiprianA.Sporea:2019ary, Panotopoulos:2018pvu}, i.e., for very high or low frequencies of the field wave modes. On the other hand, there are methods that deal with taking extremal limit to evaluate these quantities, see \cite{Cvetic:1997uw, Cvetic:2009jn, Li:2009zzf}, or analytically estimating the bounds on these greybody factors, see \cite{Boonserm:2008zg, Ngampitipan:2012dq, Ngampitipan:2013sf, Boonserm:2013dua, Boonserm:2014rma, Boonserm:2014fja, Boonserm:2017qcq}. These bounds have the advantage of being predicted in all frequency regimes, including the intermediate frequency regimes, and also for all values of the angular momentum quantum number. In particular, we are going to consider a massless minimally coupled scalar field in the radially deformed BH spacetime, and estimate these bounds to study the spectrum of the Hawking effect with the greybody factors. Especially our motivation is understanding the changes caused by the inclusion of the radial deformation parameter in a stationary and rotating BH spacetime.

In \ref{sec:horizon-structure-DK}, we begin by providing a detailed investigation of the horizon structure in the radially deformed BH spacetime. In \ref{sec:scalar-field-EOM}, we consider a massless minimally coupled scalar field in this radially deformed BH spacetime, and obtain the scalar field equation of motion. Decomposition of the scalar field in terms of the spheriodal harmonics provides one with a Schr\"odinger wave like equation, namely the \emph{Teukolsky equation} for stationary Kerr BHs, in terms of the radial tortiose coordinate. In particular, from this equation the structure of the effective potential can be perceived. Subsequently, in \ref{sec:Hawking-effect}, a study of the Hawking effect and the corresponding temperature and spectrum of the Hawking quanta are provided. Furthermore, in \ref{sec:GBfactor}, we study the bounds on the greybody factors in these radially deformed BH spacetimes considering the effective potential from \ref{sec:scalar-field-EOM}. We conclude our analysis with a discussion in \ref{sec:conclusion}.
\section{Horizon structure of a radially deformed spacetime}\label{sec:horizon-structure-DK}
%
We should mention that there are a few well-known 
deformations of the Kerr metric which serve specific purposes 
\cite{Johannsen:2011dh, Rezzolla:2014mua,Johannsen:2015hib,Konoplya:2016jvv,
Glampedakis:2017dvb}. For example, the metric provided by Johannsen and Psaltis 
is a Kerr-like metric that provides a regular spacetime everywhere outside of 
the event horizon \cite{Johannsen:2011dh}. In this case, the deformation is a 
function of both the radial coordinate $r$ and the angular coordinate $\theta$, 
and the spacetime is asymptotically the same as the Kerr spacetime. The field 
equation is not separable in this background in terms of the radial and angular 
coordinates. However, it is imperative to understand quantum field theory in the 
background black hole spacetime to realize the Hawking effect properly. That 
becomes possible in the spacetime provided by Konoplya and Zhidenko 
\cite{Konoplya:2016pmh}, which only has radial coordinate in deformation, 
enabling one to separate the field equation. This motivated us to consider this 
second type of description for the deformed Kerr black hole spacetime 
\cite{Konoplya:2016pmh} to study the Hawking effect.

We start with the following Kerr metric written in a more generic form  \cite{Konoplya:2016pmh}:
\begin{eqnarray}\label{eq:Kerr-metric}
 ds^2 &=& -\frac{N^2(r,\theta)-W^2(r,\theta)\sin^2{\theta}}{K^2(r,\theta)} dt^2 \nonumber \\
& & -2r W(r,\theta)\sin^2{\theta} dt d\phi  +r^2K^2(r,\theta)\sin^2{\theta}d\phi^2 + \nonumber\\ &&\sigma(r,\theta)\left(\frac{B^2(r,\theta)}{N^2(r,\theta)} 
dr^2+r^2d\theta^2\right)~, 
\label{eq:metric_func}
\end{eqnarray}
with,
\begin{eqnarray}\label{eq:Kerr-metric-coeff}
 N^2(r,\theta) &=& \frac{\Delta}{r^2},~ K^2(r,\theta) =
\frac{\Sigma}{r^4 \sigma(r,\theta)}~,\nonumber \\
W(r,\theta) &=& a\rs(r^2+a^2\cos^2{\theta})^{-1}\nonumber, \\
B(r,\theta) &=& 1,~ \Sigma = (r^2+a^2)^2-\Delta~ a^2 \sin^2\theta~, \nonumber \\
 \sigma(r,\theta) &=& r^{-2}(r^2+a^2\cos^2{\theta}),~\Delta=r^2-r \rs+a^2.
\end{eqnarray}
In the above expressions, $\rs=2M$, where $M$ denotes the mass of the BH, and $a$ is the angular momentum per unit mass. In order to inject the radial deformation, we use the substitution $\rs \rightarrow \rs+{\eta}/{(r^{2})}$. This substitution would not change any of the built-in properties of the spacetime including the separability condition of the Klein-Gordon equation. The only difference that distinguishes the deformed spacetime from the Kerr metric of \ref{eq:Kerr-metric}, is in $\Delta$, which now becomes, $\bar{\Delta}=r^2-r\rs+a^2-\eta/r$, and also $\Sigma$ changes to $\bar{\Sigma}$ with $\Delta$ is replaced by $\bar{\Delta}$ in the expressions of \ref{eq:Kerr-metric-coeff}.

Due to the presence of deformation parameter $\eta$, the locations of the horizons, given as $N^2 (r,\theta) = 0$, would differ from the usual Kerr case. Moreover, as $\eta$ is clearly coupled with $r$, new solutions may also emerge. To be specific, the locations of the horizons are obtained from,
\begin{equation}
r^3-r^2\rs+a^2r-\eta=0, 
\label{eq:Horizon_Deformed}
\end{equation}
and we will apply Descartes' sign rule to estimate the number of solution(s). Note that with $\eta<0$, there is no positive solution for the above case, and the naked singularity always exists. With $\eta>0$, \ref{eq:Horizon_Deformed} can either have one or three positive solution(s). In the later case, the event horizon, $\rh$ and these inner horizons, $r_1$ and $r_2$ (assume $r_1<r_2$), can be expressed in terms of the other BH parameters as
\begin{eqnarray}\label{eq:horizons-DK}
 \rh &=& \frac{\rs}{3}-\frac{2^{\frac{1}{3}}(3a^2-\rs^2)}{3\mathcal{A}^{1/3}}+\frac{\mathcal{A}^{1/3}}{3\times2^{\frac{1}{3}}}~,\nonumber\\
 r_{1} &=& \frac{\rs}{3}+\frac{(1+i\sqrt{3})(3a^2-\rs^2)}{3\times2^{\frac{2}{3}}\mathcal{A}^{1/3}}-\frac{(1-i\sqrt{3})\mathcal{A}^{1/3}}{6\times2^{\frac{1}{3}}}~,\nonumber\\
 r_{2} &=& \frac{\rs}{3}+\frac{(1-i\sqrt{3})(3a^2-\rs^2)}{3\times2^{\frac{2}{3}}\mathcal{A}^{1/3}}-\frac{(1+i\sqrt{3})\mathcal{A}^{1/3}}{6\times2^{\frac{1}{3}}}~,
 \label{eq:horizon} 
\end{eqnarray}
where, $\mathcal{A}=\beta_{1}+3\sqrt{3}~\beta_{2}$, with the expression of $\beta_{1}$ and $\beta_{2}$ given by $\beta_{1}=-9a^2 \rs+2\rs^3+27\eta$ and $\beta_{2} = \left(4a^6 -a^4\rs^2 -18a^2\rs\eta +4\rs^3\eta +27\eta^2\right)^{1/2}$. 

We may now employ the above expressions to have a deeper understanding about the horizon structure in presence of $\eta$. Based on the properties of $\mathcal{A}$, whether it is positive, negative or complex, we encounter different outcomes. In case of (a) $\mathcal{A}>0$, $\rh$ is always positive and describe the event horizon, while $r_1$ and $r_2$ are complex conjugate to each others. For (b) $\mathcal{A}<0$, say $\mathcal{A}=-\alpha$ where $\alpha>0$, \ref{eq:horizon} becomes:
\begin{eqnarray}\label{eq:horizons-DK2}
 \rh &=& \frac{\rs}{3}-\frac{2^{1/3}\exp(-i \pi/3)(3a^2-\rs^2)}{3\alpha^{1/3}}+\frac{\alpha^{1/3}\exp(i \pi/3)}{3\times 2^{1/3}},\nonumber\\
 r_{1} &=& \frac{\rs}{3}+\frac{2^{1/3}(3a^2-\rs^2)}{3\times \alpha^{1/3}}-\dfrac{\alpha^{1/3}}{3 \times 2^{1/3}},\nonumber\\
 r_{2} &=& \frac{\rs}{3}-\frac{2^{1/3}\exp(i \pi/3)(3a^2-\rs^2)}{3\alpha^{1/3}}+\frac{\alpha^{1/3}\exp(-i \pi/3)}{3\times 2^{1/3}}, \nonumber
 \\
\end{eqnarray}
which now readily gives $\rh$ and $r_{2}$ are complex conjugate to each other, and $r_1$ now becomes the outer horizon. Interestingly, if we use the $\eta=0$ limit, and the extremality condition $a=\rs/2$, we gather $\alpha=-\rs^3/4$. The above equations now gives $\rh=r_2=\rs/2$ and $r_1=0$, which is the usual Kerr case. This serves as an useful validation of our solution. Now we consider the last case, i.e., (c) $\mathcal{A}=\beta_1+i 3\sqrt{3}\sqrt{-\beta_2^2}$. In this case, we further simplify $\mathcal{A}$ as, $\mathcal{A}=2(-3 a^2+\rs^2)^{3/2} \exp(i\alpha)$, where 
\begin{equation}
\scalebox{0.9}{$\tan\alpha=\dfrac{\left\{3\sqrt{3}(-4 a^6+a^4 \rs^2+18 a^2 \rs \eta-4 \eta \rs^3-27 \eta^2)\right\}^{1/2}}{\left(-9a^2\rs+2 \rs^3+27 \eta\right)}.$}
\end{equation}
As it can be shown that in this case, the imaginary part of all of the above expressions would identically vanish, and $\rh$ continue to be the event horizon.

\begin{figure}[htp]
\includegraphics[height=5.4cm,width=0.85\linewidth]{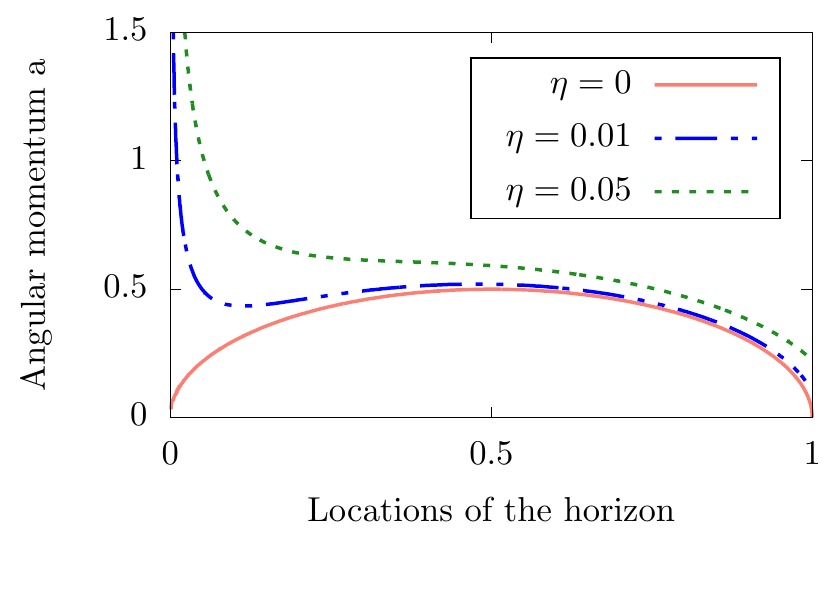}
\caption{In the above figure, we demonstrate the contours of the horizon structure in a radially deformed spacetime. Depending on the deformation parameter $\eta$, the plots change accordingly and can have maximum three positive real solutions. Beside the inner and outer horizon, the presence of $\eta$ introduces another real root which always exists. This would indicate that for no ranges of BH parameters the singularity can be naked. Here we set $\rs=1$.}
\label{Figure_01}
\end{figure}

To summarize, the presence of $\eta$ manifests an additional horizon other than the event and Cauchy horizon. Except for the $\mathcal{A}<0$ (note that the extremal Kerr is a special case in this limit where the imaginary parts of $\rh$ and $r_2$ become zero), $\rh$ continue to be the event horizon, where $r_1$ and $r_2$ are the inner horizons. For a clear exposition of this horizon structure in presence of $\eta$, we illustrate the horizon structure in \ref{Figure_01}. From this figure, we observe that for a specific $\eta$, at the same $a$, there can be different numbers of horizons or real roots to the solution of $\bar{\Delta}$. For example, for $\eta=0$, denoted by the solid red line and signifying zero deformation, there are two roots, and these two roots merge to a single one at the point $a=0.5=r_{s}/2$ in the extremal case. On the other hand, when $\eta=0.01$ or $\eta=0.05$, denoted by the dot-dashed blue and dotted green lines respectively, there can be three, two, or even one real roots to the solution for a fixed value of $a$.

The other intrinsic property that is associated with a rotating spacetime is known as the ergoregion, where no  observer can be kept stand still. This is given by the condition $g_{tt}=0$, which according to \ref{eq:metric_func} becomes $N^2=W^2 \sin^2\theta$. Ergoregion is closely related with frame dragging and \emph{zero angular momentum observer (ZAMO)}, both are well known relativistic effects \cite{wald2010general}. Given that ZAMO is relevant for the present purpose, we shall briefly discuss the same as follows. For the present spacetime, the angular velocity of ZAMO at a given $(r,\theta)$ is 
\begin{eqnarray}\label{eq:AngularVel-ZAMO}
 \Omega &=& \frac{g^{t\phi}}{g^{tt}} = \frac{a(r^2\rs+\eta)}{r\bar{\Sigma}}~.
\end{eqnarray}
If one considers the maximally extended case of $\theta=0$ or $\theta=\pi$ for the expression of this angular velocity, then
\begin{eqnarray}\label{eq:AngularVel-ZAMOm}
 \Omega = \Omega_{m}(r) = \frac{a}{(r^2+a^2)^2} \left(r^2+a^2-\bar{\Delta}\right)~,
\end{eqnarray}
which becomes $\Omega_{H}=a/(\rh^2+a^2)$, on the event horizon. 

\begin{figure}[htp]
\includegraphics[height=5.3cm,width=0.85\linewidth]{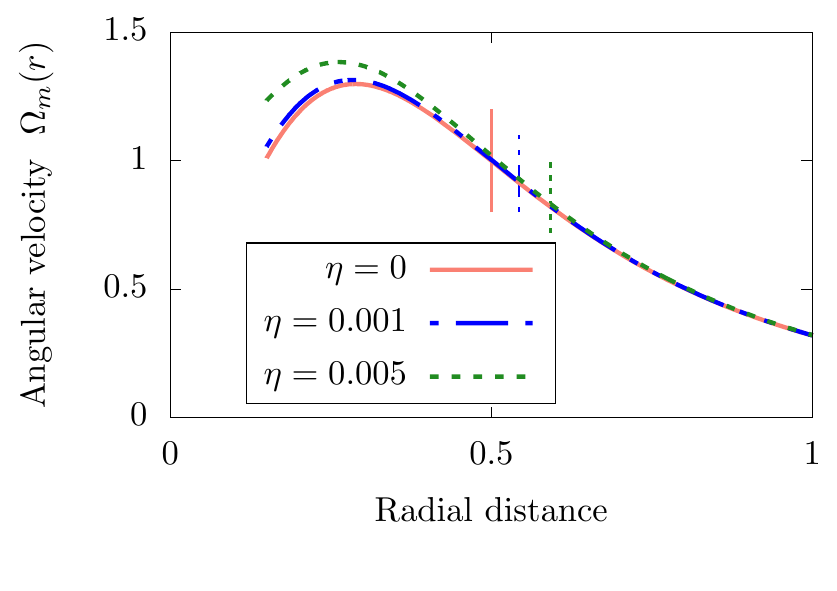}
\caption{The above figure demonstrates the variation of the angular velocity (given by the curves) with the radial distance, and the vertical lines denote the locations of the event horizons. For different values of $\eta$, it can be shown that the peaks of $\Omega_{m}$ always reside inside the event horizon. Therefore, outside the horizon, angular velocity of ZAMO is a monotonically decreasing function of $r$. We take, $a=\rs/2=0.5$.}
\label{Figure_02}
\end{figure}

Now we move on to understand the nature of $\Omega_{m}$ as a function of $r$, which we will see to be appearing in the scalar field equation of motion in the next section. In order to prove that $\Omega_{m}$ is a monotonic function in $r$ outside the event horizon, which is often required in the context of greybody factor, we may recall the case with $\eta=0$ first. It can be shown that in the Kerr case, $\Omega_{m}$ has a peak at $r=a/\sqrt{3}$, which always lies inside the outer event horizon and henceforth, the function is monotonically decreasing outside the horizon. Unfortunately, with the $\eta$ being present, the analysis becomes more involved and a straightforward solution to $r$ turns out to be unlikely. However, we carry out an approximate analysis to incorporate the effects of $\eta$ up to the linear order terms. We note that the peak of $\Omega_{m}$ as $r$ varies, located at $r_{\rm peak}$, has now become
\begin{equation}\label{eq:rpeak-Omega}
r_{\rm peak}=a/\sqrt{3}-\dfrac{4a \eta}{\sqrt{3}(a^2 \rs+5\eta)}+\mathcal{O}(\eta)^2,
\end{equation}
which shifts close to the singularity compared to the Kerr case. Similarly, we can find that the expression of the location of outer horizon changes as:
\begin{eqnarray}\label{eq:rpeak-Omega2}
\rh \vert_{\eta \ll 1}=\frac{\rs}{2}+\frac{\sqrt{\rs^2-4a^2}}{2}+\Bigl(\eta/\sqrt{\rs^2-4a^2}\Bigr)\Bigl(\rs+\nonumber \\
\sqrt{\rs^2-4a^2}\Bigr)+\mathcal{O}(\eta^2).  \nonumber \\   
\end{eqnarray}
The above equation clearly states that the outer horizon shifts away from the singularity. Therefore, we conclude that the addition of $\eta$ will shift the outer horizon away from singularity, and move the angular velocity's peak close to the singularity. This essentially assures that outside the outer horizon, $\Omega_{m}$ is montonic in $r$. For an illustration of this incident, we plot $\Omega_{m}$ for various values of $\eta$ in \ref{Figure_02}. 
As it can be realized, with $\eta \neq 0$, say $\eta=10^{-2}$, the maxima in $\Omega_{m}$ is covered within the outer event horizon. Larger the value of $\eta$, the peak gets shifted close to the singularity, and the outer horizon shifts away from singularity. This indicates that for any nonzero value of $\eta$, the monotonicity of $\Omega_{m}$ is retained outside the event horizon. 
\section{The scalar field equation of motion}\label{sec:scalar-field-EOM}
\begin{figure}[htp]
\includegraphics[height=5cm,width=0.85\linewidth]{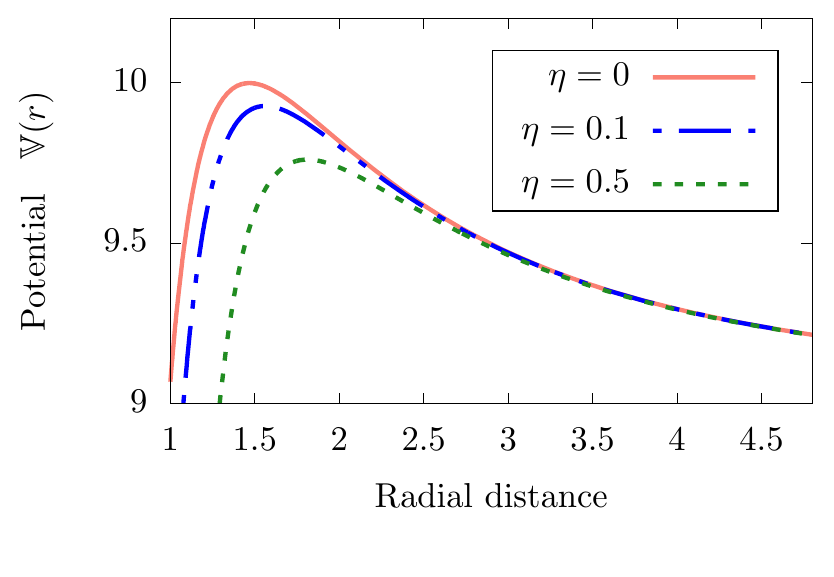}
\caption{The above figure illustrates the scalar field potential $\mathbb{V}(r)$ from \ref{eq:potential-scalar-DK} for different values of $\eta$. We set the other parameters at $a=0.1$, $l=2$, $\omega=3$, $m=0$, $\rs=1$.}
\label{Figure_03}
\end{figure}

In this deformed spacetime we consider a massless minimally coupled free scalar field $\Phi(x)$ described by the action
\begin{equation}
    \mathcal{S}_{\Phi}=\int d^4x \left[-\frac{1}{2}\sqrt{-g}g^{\mu \nu}\partial_{\mu}\Phi(x) \partial_{\nu}\Phi(x) \right]~.
\end{equation}
The variation of this action with respect to the scalar field $\Phi$ provides one with the scalar filed equation of motion
\begin{equation}
   \Box\Phi(x) \frac{1}{\sqrt{-g}}\partial_{\mu }\Bigl(\sqrt{-g}g^{\mu \nu}\partial_{\nu} \Phi(x)\Bigr)=0.\label{eq:field-EOM}
\end{equation}
By substituting the metric components from \ref{eq:Kerr-metric} and \ref{eq:Kerr-metric-coeff} with the expressions of $\bar{\Delta}$ and $\bar{\Sigma}$ corresponding to the deformed spacetime, the equation of motion becomes
\begin{eqnarray}\label{eq:field-EOM1}
  && \scalebox{0.97}{$-\dfrac{\bar{\Sigma} \sin\theta}{\bar{\Delta}}\partial^2_{t} \Phi+\dfrac{\bar{\Delta}-a^2 \sin^2\theta}{\bar{\Delta}\sin\theta}\partial^2_{\phi} \Phi -\dfrac{2a\sin\theta}{\bar{\Delta}}\left(r^2+a^2-\bar{\Delta}\right)$} \nonumber \\
 ~&&~~~~~ \partial_{t}\partial_{\phi}\Phi+\partial_{r} \left(\bar{\Delta}\sin\theta \partial_r \Phi\right)+\partial_{\theta} \left(\sin\theta \partial_{\theta} \Phi\right)=0~. 
\end{eqnarray}
Like the Kerr BH here also the metric components are independent of time $t$ and azimuthal angle $\phi$, which suggests a field decomposition of the form $\Phi(t,r,\theta,\phi)=\exp(-i \omega t+i m \phi) R(r) S(\theta) /\sqrt{r^2+a^2}$, where $S(\theta)$ denotes spheroidal harmonics. Then using this field decomposition and the tortoise coordinate $\rstar$, defined from the expression 
\begin{equation}\label{eq:tortoise-coord-DK}
 d\rstar = \frac{r^2+a^2}{\bar{\bigtriangleup}} dr~,
\end{equation}
the previous scalar field equation of motion from \ref{eq:field-EOM1} can be expressed as
\begin{eqnarray}\label{eq:radial-eom-DK}
    \partial^2_{r_*}R+\Bigl[\omega^2-\mathbb{V}(r)\Bigr]R=0~.
\end{eqnarray}
In this deformed geometry, one may express the effective potential $\mathbb{V}(r)$ as 
\begin{eqnarray}\label{eq:potential-scalar-DK}
  \mathbb{V}(r)=\dfrac{\bar{\Delta}}{(r^2+a^2)^2}\Bigl\{A^{\omega}_{lm}+\dfrac{(r\bar{\Delta})^{\prime}}{r^2+a^2}-\dfrac{3 \bar{\Delta}r^2}{(r^2+a^2)^2}\Bigr\} \nonumber \\
   -\dfrac{a^2 m^2}{(r^2+a^2)^2}+\dfrac{2ma\omega}{(r^2+a^2)^2}(r^2+a^2-\bar{\Delta}), 
   \label{potential_exact}
\end{eqnarray}
where, $A^{\omega}_{lm}$ denotes the eigenvalue corresponding to the spheroidal harmonics equation. In slow rotation limit this eigenvalue can be expressed as
\begin{equation}\label{eq:Spheriodal-Harmonics-eig}
A^{\omega}_{lm} = l(l+1) - 2m~a\omega +\mathcal{O}\left[\left(a\omega\right)^2\right]~.
\end{equation}
This Schr\"odinger wave like equation of \ref{eq:radial-eom-DK} in deformed BH spacetime resembles the Teukolsky equation \cite{Kokkotas:2010zd} from general Kerr BH spacetime. It should be noted that the from this equation the Regge-Wheeler \cite{Fiziev:2005ki, Boonserm:2013dua} like equation corresponding to a deformed Schwarzschild spacetime can be obtained quite easily making $a=0$. In this non-rotating limit, i.e., $a=0$, the expression of potential $\mathbb{V}(r)$ from \ref{eq:potential-scalar-DK} becomes
\begin{equation}
  \mathbb{V}(r)=  (1-\rs/r-\eta/r^3)\left\{\dfrac{l(l+1)}{r^2}+\dfrac{\rs}{r^3}+\dfrac{3\eta}{r^5}\right\}.
\end{equation}
Here we have used the fact that when $a=0$ the eigenvalue $A^{\omega}_{lm}$ of spheroidal harmonics equation signifies spherical harmonics with expression $A_{lm}=l(l+1)$. In the geometrical optics limit, i.e., $l \gg 1$, we gather that the above equation encounters an maxima at the photon orbit, $r_{\rm ph}$, which is a solution of the equation:
\begin{eqnarray}
2 r^3_{\rm ph}-3\rs r^2_{\rm ph}-5\eta=0.
\end{eqnarray}
With $\eta=0$, we arrive ar $r_{\rm ph}=3\rs/2$, which is the location of photon orbit in \S~BH \cite{Berti:2014bla, Munoz:2014}. In \ref{Figure_03}, we depict the potential $\mathbb{V}(r)$ from \ref{eq:potential-scalar-DK} for different values of the parameter $\eta$. As it can be evident that there exist a maxima, $r_{\rm peak}$, in each of these figures, which shifts based on the values of $\eta$.

In passing, we should also touch upon an important consequence of the above potential. It can be sensed from \ref{Figure_03} that in either side of the peak (at $r={r}_{\rm peak}$), the function $\mathbb{V}(r)$ is monotonic. In particular, from the event horizon $\rh$ to $r_{\rm peak}$, it is monotonically increasing, and from $r_{\rm peak}$ to $\infty$, it is monotonically decreasing. It is similar to saying that there will be a single peak outside the event horizon. In the presence of both $a$ and $\eta$, the concern may arise weather this feature remains intact or not. Consequently, even with $\eta$ set to zero, it is unlikely to guess from \ref{potential_exact} that the potential will have a single peak outside the event horizon. Therefore, an analytical proof is beyond expectation. What we found that within the weak rotation approximation ($a \omega \ll 1$) \cite{Boonserm:2014rma}, this property is always valid for a wide range of various BH parameters, and different modes.
\section{Hawking effect in radially deformed spacetime}\label{sec:Hawking-effect}
In the original work \cite{hawking1975}, the thermal nature of the Hawking effect is realized through the usage of Bogoliubov transformation between the ingoing and outgoing field modes described in terms of the null coordinates. The Hawking effect can be realized through other various means, like using tunnelling formalism \cite{Parikh:1999mf, Angheben:2005rm, Banerjee:2008cf, Umetsu:2009HRKN, Yale:2010tn, Vanzo:2011wq, Feng:2015exa}, path integral approach \cite{Hartle:1976tp}, conformal symmetry \cite{Agullo:2010hi}, via anomalies \cite{Murata:2006pt, Iso:2006ut, Jiang:2007wj}, canonical formulation \cite{Barman:2017fzh, Barman:2018ina, Hossain:2019bmy, Barman:2019vst}, and as an effect of near horizon local instability \cite{Dalui:2019esx, Dalui:2020qpt, Majhi:2021bwo}. However, the conclusion remains the same, i.e., an asymptotic observer will perceive the BH horizon with some temperature proportional to its surface gravity. In particular, in the case of a deformed BH spacetime, the number density of the Hawking quanta perceived by an asymptotic observer with frequency $\omega$ and angular momentum quantum number $m$ will be given by
\begin{equation}\label{eq:Hawking-NumberDensity}
    N_{\omega} = \frac{\Gamma(\omega)}{e^{2\pi(\omega-m\Omega_{H})/\kappa_{H}}-1}~,
\end{equation}
where, $\kappa_{H}$, $\Omega_{H}$, and $\Gamma(\omega)$ denote the surface gravity, angular velocity at the event horizon, and the greybody factor respectively. For modes with $m\Omega_{H}>\omega$ this expression gives rise to the so called super-radiance phenomenon \cite{book:PadmanabhanGrav}. From the Planckian distribution of \ref{eq:Hawking-NumberDensity} the characteristic temperature corresponding to the Hawking effect is $T_{H}=\kappa_{H}/2\pi$. In a radially deformed BH spacetime (with $\rs\to\rs+\eta/r^2$ deformation) the surface gravity at the outer horizon $r=\rh$, which gives the temperature of the horizon, can be found out to be 
\begin{equation}\label{eq:surface-Grav-DK}
 \kappa_{_{H}} = \frac{(\rh-r_{1})(\rh-r_{2})}{2\rh(\rh^2+a^2)}~,
\end{equation}
the evaluation of which is given in \ref{Apn:SurfaceGrav-OutHorizon}. The inner horizons, $r_{1}$ and $r_{2}$  can exist if they are real or cease to exist if they become imaginary for certain values of $\eta$. From \ref{eq:horizons-DK}, \ref{eq:horizons-DK2} it is clear that the position of the horizon now has a signature of the deformation which will also be apparent in the spectrum of the Hawking effect. Another quantity is the angular velocity at the outer horizon from \ref{eq:AngularVel-ZAMO}, which will also carry the signature of the deformation in the spectrum of \ref{eq:Hawking-NumberDensity}. In our following discussion we study the nature of surface gravity and angular velocity at the outer horizon, in particular, we observe how it changes with varying $\eta$.
\begin{figure}[htp]
\includegraphics[height=5cm,width=0.85\linewidth]{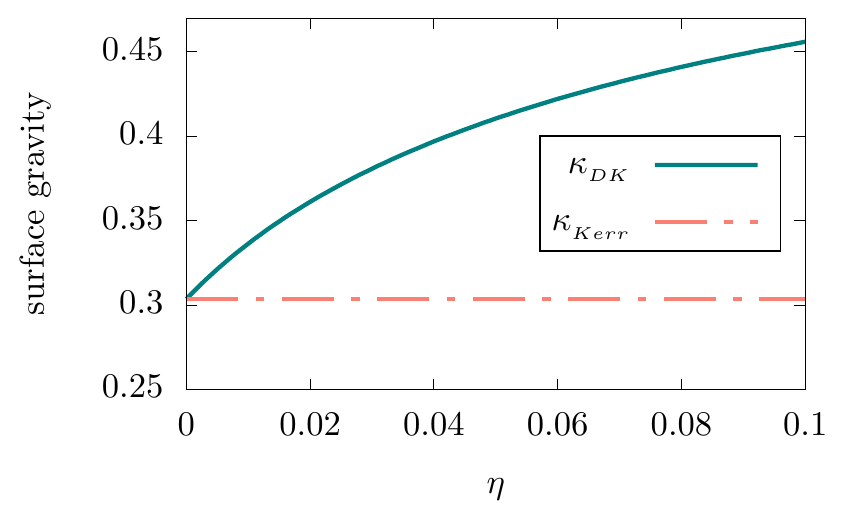}
\caption{In the above figure the event horizon's surface gravity $\kappa_{_{H}}$ in a deformed BH spacetime, denoted by $\kappa_{_{DK}}$, is plotted with respect to $\eta$. We have set the other parameters at $a=0.45$, $\rs=1$. The plot also depicts the case when $\eta=0$, i.e., the Kerr BH scenario, denoted by $\kappa_{_{Kerr}}$.}
\label{fig:SG-etaV}
\end{figure}
\begin{figure}[htp]
\includegraphics[height=5cm,width=0.85\linewidth]{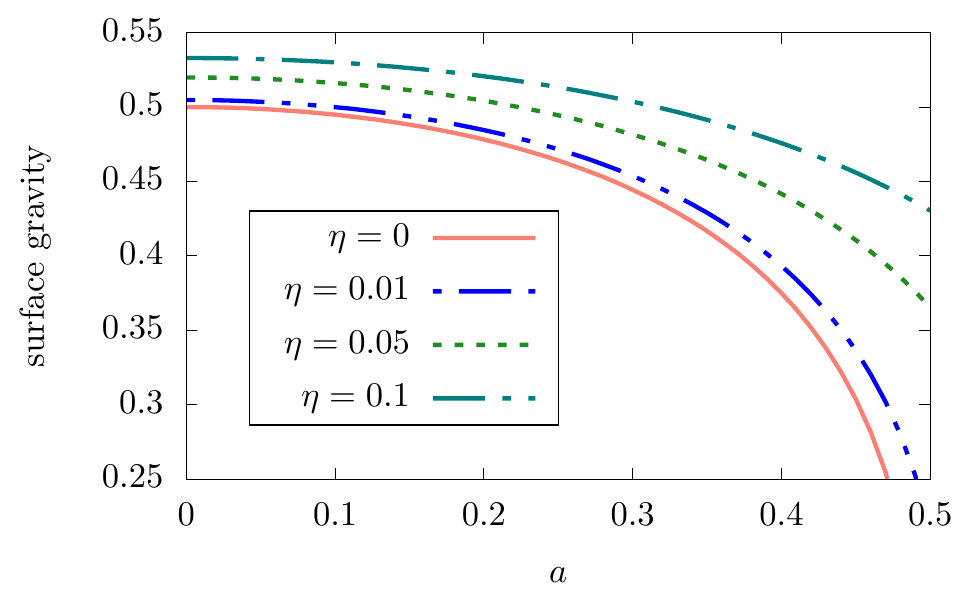}
\caption{In the above figure the surface gravity in a deformed spacetime is plotted with respect to varying $a$ with different $\eta$. We have set the other parameter at $\rs=1$.}
\label{fig:SG-aV-etaD}
\end{figure}

In \ref{fig:SG-etaV} we have plotted the surface gravity corresponding to the outer horizon $r=\rh$ in a deformed spacetime with respect to varying $\eta$. The figure is obtained considering the other BH parameters to be $\rs=1$ and $a=0.45$. In the same figure the zero $\eta$ situation is also depicted by a dash-dotted line. From this figure it can be observed that as the value of the deformation $\eta$ increases the value of the horizon's surface gravity also increases, signifying an increase in the characteristic temperature of the Hawking effect.

In \ref{fig:SG-aV-etaD}, we have plotted the event horizon's surface gravity, which gives the Hawking temperature, with respect to varying angular momentum parameter $a$ for different values of the deformation parameter $\eta$. From this figure, we note that as the value of $\eta$ increases, surface gravity departs further from the Kerr case and also increases corresponding to different fixed values of $a$.

\begin{figure}[htp]
\includegraphics[height=5cm,width=0.85\linewidth]{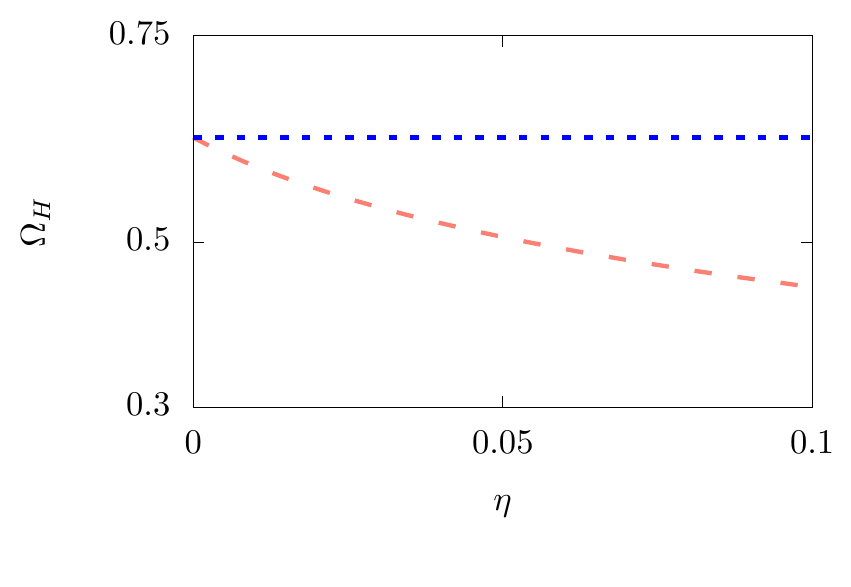}
\caption{Angular velocity of the horizon is shown for $a=0.45$, and $\rs=1$. We retrieve the kerr case for $\eta=0$ limit, which is shown as the curve constant along $x$-axis.
}
\label{fig:OEH-etaV}
\end{figure}

Similarly, in \ref{fig:OEH-etaV}, we demonstrate how the angular velocity of the event horizon changes with deformation parameter. Note that $\Omega_{H}$ is implicitly affected by $\eta$ through the value of $\rh$. 

\section{The bounds on the greybody factor}\label{sec:GBfactor}
The spectrum of the Hawking effect \ref{eq:Hawking-NumberDensity} is given by a greybody distribution as perceived by an asymptotic future observer. The greybody factor arises from the transmission amplitude of the field modes through the effective potential outside the horizon, for the modes nearly escaping the formation of the horizon and travelling from the near-horizon region to an asymptotic observer. In this section we estimate the bounds on the greybody factor, see \cite{Boonserm:2008zg, Ngampitipan:2013sf, Boonserm:2014rma}, which can be analytically expressed in all frequency range, for certain general field momenta and general spacetime dimensionalities. These bounds on the greybody factor can be expressed as
\begin{equation}\label{eq:bound-GBF}
 \Gamma(\omega) \geq 
\sech^2{\Theta}~,
\end{equation}
where
\begin{equation}\label{eq:theta-GBF}
 \Theta = \int_{-\infty}^{\infty}\frac{\sqrt{(h')^2+(\omega^2-\mathbb{V}(r)-h^2)^2}}{2h} d\rstar ~.
\end{equation}
Here, $\mathbb{V}(r)$ denotes the effective potential corresponding to a massless minimally coupled scalar field and $\omega$ corresponds to the frequency of field mode. Furthermore, $h\equiv h(\rstar)$ is some positive function satisfying the condition $h(-\infty) =h(\infty) =\omega$. The tortoise coordinate $\rstar$ is obtained from the expression of \ref{eq:tortoise-coord-DK}. From \ref{eq:potential-scalar-DK} one can obtain the effective potential correspond to a massless minimally coupled free scalar field in a deformed BH spacetime and express $\omega^2-\mathbb{V}(r)$ in another convenient form as 
\begin{eqnarray}\label{eq:potential-scalar-DK2}
 \scalebox{0.97}{$\omega^2-\mathbb{V}(r)$} &=& \scalebox{0.97}{$(\omega-m\Omega_{m})^2+\mathbb{U}(r)$}\nonumber\\
~ &=& \scalebox{0.97}{$(\omega-m\Omega_{m})^2+\frac{m^2 a^2 
\bar{\bigtriangleup}}{(r^2+a^2)^2}$} 
\left(\bar{\bigtriangleup}+\frac{\eta}{r^{n+1}}+2r r_{s}\right)\nonumber\\
~&& \scalebox{0.97}{$-\frac{\bar{\bigtriangleup}}{(r^2+a^2)^2} \left\{ \mathcal{A}_{lm}^{\omega} 
+ \frac{(r\bar{\bigtriangleup})'(r^2+a^2)-3\bar{\bigtriangleup}r^2}{(r^2+a^2)^2} 
\right\}$}~.
\end{eqnarray}\vspace{0.05cm}

Now one can consider the simplest choice of the positive function $h(\rstar)=\omega$, as also done in \cite{Boonserm:2008zg} for Schwarzschild BHs. However, it will only be fruitful for the case of $m=0$, as the first quantity from \ref{eq:potential-scalar-DK2} contributes to a diverging term in the integration of \ref{eq:theta-GBF} for the calculation of the bound for $m\neq0$. We then consider the evaluation of these bounds on the greybody factor in a case by case manner as done in \cite{Boonserm:2014rma}. In particular, in separate situations for the angular momentum quantum number $m=0$ and $m\neq 0$.
\subsection{The case of m=0 :} \label{sec:m_0_case}
This particular case of $m=0$ is the simplest and provides an overall picture of the bounds on the greybody factor. In this case we are going to consider the positive function to be $h=\omega$, i.e., $h'=0$. Now as one makes the choice of $m=0$ in the expression of the scalar field effective potential from \ref{eq:potential-scalar-DK2}, one gets
\begin{equation}\label{eq:potential-scalar-DK-m0}
 \omega^2-\mathbb{V} = \omega^2
 -\frac{\bar{\bigtriangleup}}{(r^2+a^2)^2} \left[ \mathcal{A}_{l0}^{\omega} 
+ \frac{(r\bar{\bigtriangleup})'(r^2+a^2)-3\bar{\bigtriangleup}r^2}{(r^2+a^2)^2} 
\right].
\end{equation}
Then with this expression of the potential and the above choice of the positive function $h$ one can represent the bound on the greybody factor to be 
\begin{equation}\label{eq:GBF-bound-m0-1}
 \Gamma(\omega) \geq 
\sech^2{\left(\frac{\mathscr{I}^{\omega}_{l0}}{2\omega}\right)}~,
\end{equation}
where, $\mathscr{I}^{\omega}_{l0}=\mathscr{I}^{\omega}_{lm}(m=0)$ and the quantity $\mathscr{I}^{\omega}_{lm}$ in general is defined by the integral
\begin{equation}\label{eq:bound-GBF-m0}
\mathscr{I}^{\omega}_{lm} = \int_{\rh}^{\infty} dr \left[ \frac{\mathcal{A}_{lm}^{\omega}}{r^2+a^2} 
+ \frac{(r\bar{\bigtriangleup})'(r^2+a^2)-3\bar{\bigtriangleup}r^2}{(r^2+a^2)^3} \right]~.
\end{equation}
%
\begin{figure}[htp]
\includegraphics[height=5cm,width=0.95\linewidth]{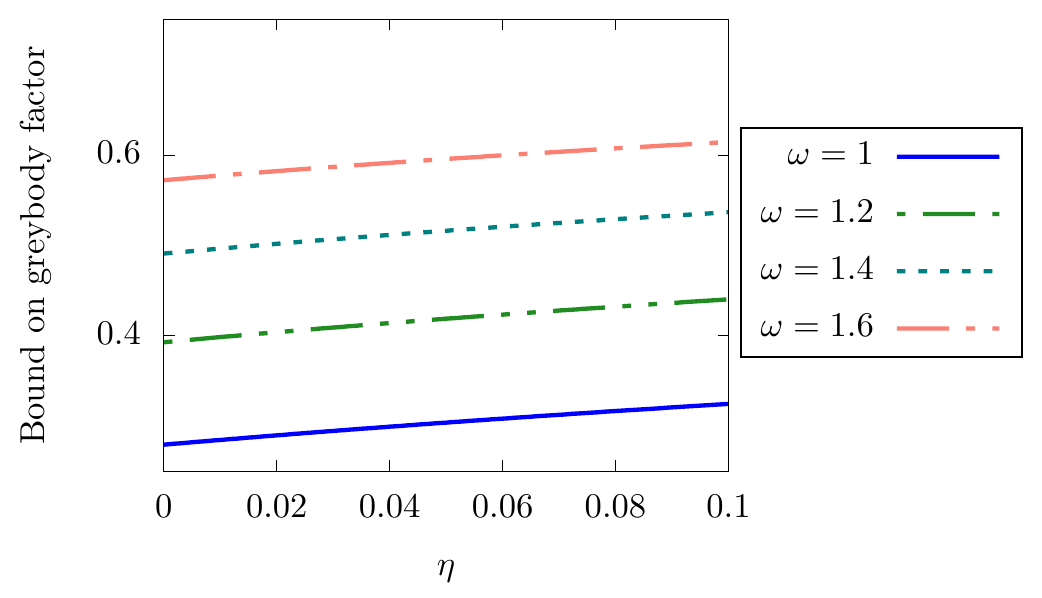}
\caption{In the above figure the lower bound on the greybody factor in a deformed spacetime for $m=0$ is plotted with respect to varying $\eta$ for different fixed $\omega$. We have set the other parameters at $l=1$, $\rs=1$, and $a=0.05$. We note that the quantity $a\omega\ll1$ so that the approximation of \ref{eq:Spheriodal-Harmonics-eig} can be made for the Spheriodal harmonics eigenvalues.}
\label{fig:BGBF-etaV}
\end{figure}
\begin{figure}[htp]
\includegraphics[height=5cm,width=0.85\linewidth]{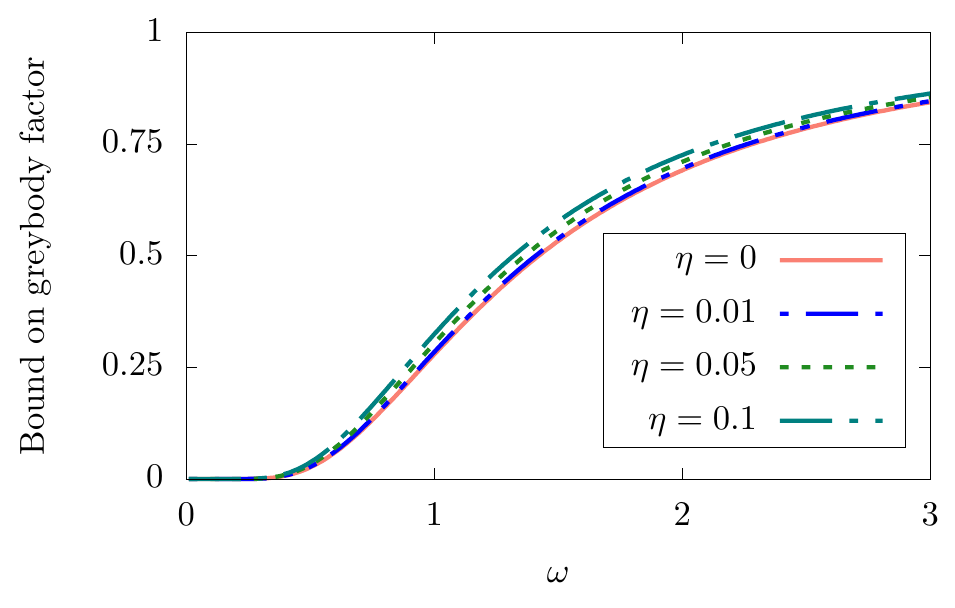}
\caption{In the above figure the lower bound on the greybody factor in a deformed BH spacetime for $m=0$ is plotted with respect to varying $\omega$ for different fixed $\eta$. We have set the other parameters at $l=1$, $\rs=1$, and $a=0.05$.}
\label{fig:BGBF-wV-etaD}
\end{figure}
One can evaluate this integral by step by step carrying out all of its components which we describe in \ref{Apn:Useful_rel}.  Utilizing these expressions, we obtain the integral of \ref{eq:bound-GBF-m0}  for the case of $m=0$ as
\begin{equation}\label{eq:bound-GBF-m0-f}
\mathscr{I}^{\omega}_{l0} = \scalebox{0.95}{$\mathcal{A}_{l0}^{\omega} \mathcal{I}_{1} +a^2
 \mathcal{I}_{2} +\rs~  \mathcal{I}_{3}^{2} -3a^2\rs~\mathcal{I}_{3}^{3}+3\eta\mathcal{I}^{0}_{4}$},
\end{equation}
where the expressions for $\mathcal{I}_{1}$, $\mathcal{I}_{2}$, $\mathcal{I}_{3}$, and $\mathcal{I}^{0}_{4}$ are given in \ref{Apn:Useful_rel}. With this expression of the integral $\mathscr{I}^{\omega}_{l0}$ from \ref{eq:bound-GBF-m0-f} and putting it in \ref{eq:GBF-bound-m0-1} one can find out the bound on the greybody factor in a deformed spacetime for $m=0$. We have further plotted this bound in \ref{fig:BGBF-etaV} for varying $\eta$ with different fixed $\omega$, and in \ref{fig:BGBF-wV-etaD} for varying $\omega$ with different fixed $\eta$. From \ref{fig:BGBF-etaV} it can be observed that as the value of the deformation parameter increases the bound on the greybody factor also increases. However, it never goes beyond the upper limit $1$. The bound also increases with increasing frequency $\omega$ of the wave mode, which can be observed from both \ref{fig:BGBF-etaV} and \ref{fig:BGBF-wV-etaD}.

\subsection{The case of $m\neq0$} \label{sec:m_neq_0_case}
As already observed from \ref{Figure_02} and the discussions related to \ref{eq:rpeak-Omega} and \ref{eq:rpeak-Omega2}, we consider the angular velocity of ZAMO to be monotonic, in particular monotonic decreasing, in the entire range between the outer horizon and the asymptotic infinity. We are going to consider this property and find out the bounds on the greybody factor. Because of this feature then one can also consider the form of the positive function $h$ in terms of $\Omega_{m}(r)$, as done in \cite{Boonserm:2014rma}, which can be made monotonic in the region between the event horizon and the asymptotic infinity. With this consideration from \ref{eq:bound-GBF} the bound on the greybody factor becomes
\begin{equation}\label{eq:bound-GBF-mn0}
 \scalebox{0.98}{$\Gamma(\omega) \geq 
\sech^2{\left\{\frac{1}{2}\int_{-\infty}^{\infty} d\rstar \frac{|h'|}{h} + \frac{1}{2}\int_{-\infty}^{\infty} d\rstar \frac{|\omega^2-\mathbb{V}-h^2|}{h} \right\}}~$},
\end{equation}
where, we have sought the help of the \emph{triangle inequality} to express the numerator of \ref{eq:theta-GBF}. This expression further simplifies to 
\begin{equation}\label{eq:bound-GBF-mn0-1}
 \scalebox{0.98}{$\Gamma(\omega) \geq 
\sech^2{\left\{\frac{1}{2} \left|\ln{\left[\frac{h(\infty)}{h(-\infty)}\right]}\right| + \frac{1}{2}\int_{-\infty}^{\infty} d\rstar \frac{|\omega^2-\mathbb{V}-h^2|}{h} \right\}}$}.
\end{equation}
We consider $\Omega_{H}\equiv\Omega(\rh)$ to be the angular velocity at the event horizon. In regard to the monotonicity of $\Omega_{m}(r)$, it should be noted that $\Omega_{m}(r)$ is always smaller than $\Omega_{H}$ in the region outside the outer horizon, which can be used for the monotonicity of $h$. Furthermore, this monotonicity can be achieved in two different regions, namely in the non super-radiant regimes $\omega> m\Omega_{H}$ or $m< \ms$ with $\ms=\omega/\Omega_{H}$, and for the case of the super-radiant modes of $m> \ms$. In the following discussion we shall consider the case with the non super-radiant and super-radiant modes separately.

\subsubsection{Non super-radiant modes $m<\ms$}

First we consider the non super-radiant modes, i.e., $m<\ms$, where we can have another two possibilities either $m<0$ or $m\in(0,\ms)$. The function $h$ in both cases are considered to be $h=\omega-m\Omega_{m}(r)$, which is positive. Let us check for the first quantity of \ref{eq:bound-GBF-mn0-1} in these two scenarios. In particular we have
\begin{eqnarray}\label{eq:I1-ineq-nonSup-GBF-mn0}
 \frac{h(\infty)}{h(-\infty)} &=& \frac{1}{1-\frac{m\Omega_{H}}{\omega}}\nonumber\\
 ~&<& 1, ~~\textup{when, }~m<0 \nonumber\\
 ~&>& 1, ~~\textup{when, }~m\in(0,\ms) ~.
\end{eqnarray}
On the other hand, using the expression of $(\omega^2-\mathbb{V}(r))$ from \ref{eq:potential-scalar-DK2} the second quantity of \ref{eq:bound-GBF-mn0-1} becomes 
\begin{eqnarray}\label{eq:I2-ineq-nonSup-GBF-mn0}
 \frac{1}{2}\int_{-\infty}^{\infty} d\rstar \tfrac{|\omega^2-\mathbb{V}(r)-h^2|}{h} &=& \frac{1}{2}\int_{-\infty}^{\infty} dr \tfrac{|\mathcal{U}(r)|}{\omega-m\Omega_{m}(r)}\nonumber\\
 ~&>& \frac{1}{2}\int_{\rh}^{\infty} dr \tfrac{|\mathcal{U}(r)|}{\omega-m\Omega_{H}}, ~\textup{when}~m<0 \nonumber\\
 ~&>& \frac{1}{2}\int_{\rh}^{\infty} dr \tfrac{|\mathcal{U}(r)|}{\omega}, ~\textup{when}~m\in(0,\ms) ,\nonumber\\
\end{eqnarray}
where, the expression $\mathcal{U}(r) = \mathbb{U}(r) (r^2+a^2)/ \bar{\Delta}$, and $\mathbb{U}(r)$ is obtained from \ref{eq:potential-scalar-DK2}. Then for the scenario $m<0$ of the non super-radiant modes $m<\ms$, the bound on the greybody factor from \ref{eq:bound-GBF-mn0-1} can be obtained using the outcomes of \ref{eq:I1-ineq-nonSup-GBF-mn0} and \ref{eq:I2-ineq-nonSup-GBF-mn0} as
\begin{eqnarray}\label{eq:bound-GBF-mn0-2}
 \Gamma(\omega) &\geq& 
\sech^2 \left\{\frac{1}{2} \ln{\left(1-\frac{m\Omega_{H}}{\omega}\right)} +\frac{\mathscr{I}^{\omega}_{lm}}{2(\omega-m\Omega_{H})}\right.\nonumber\\ 
~&& \left. ~~-~\frac{m^2a^2}{2(\omega-m\Omega_{H})} \left(\mathcal{I}_{2}+\rs\mathcal{I}_{3}^{3}+\eta\mathcal{I}^{0}_{5}\right) \right\}~.
\end{eqnarray}

\begin{figure}[htp]
\includegraphics[height=5cm,width=0.95\linewidth]{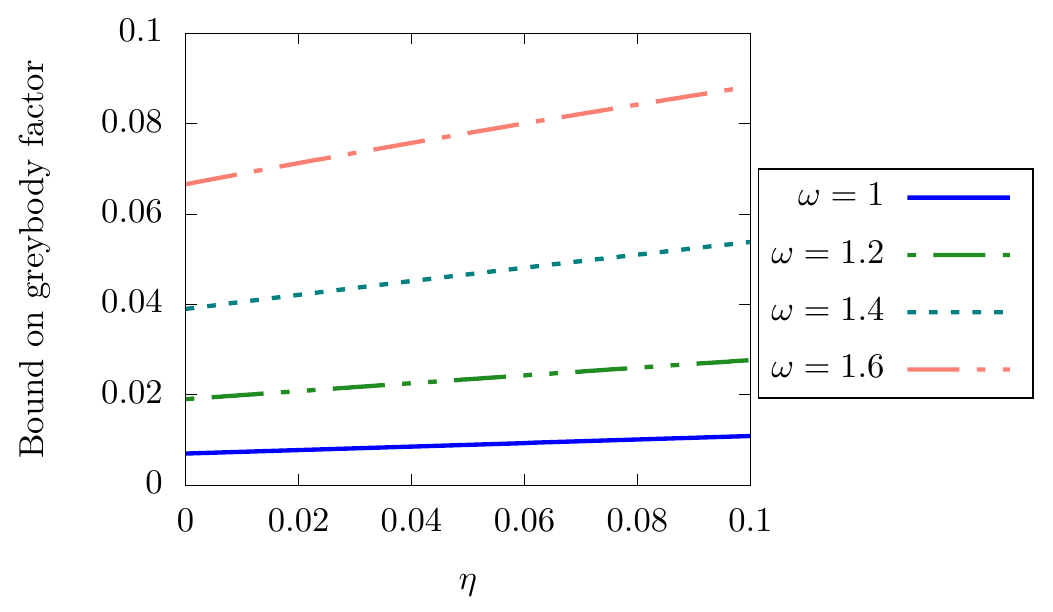}
\caption{In the above figure the lower bound on the greybody factor in a radially deformed BH spacetime is plotted with respect to varying $\eta$ for different fixed $\omega$. We have set the other parameters at $l=2$, $m=-1$, $\rs=1$, and $a=0.05$. This particular case signify one situation of $m\neq0$.}
\label{fig:BGBF-etaV-DK-mLms}
\end{figure}

On the other hand, for the scenario $m\in(0,\ms)$ of the non super-radiant modes the bound on the greybody factor from \ref{eq:bound-GBF-mn0-1} is
\begin{eqnarray}\label{eq:bound-GBF-mn0-3}
 \Gamma(\omega) &\geq& 
\sech^2 \left\{-\frac{1}{2} \ln{\left(1-\frac{m\Omega_{H}}{\omega}\right)} +\frac{\mathscr{I}^{\omega}_{lm}}{2\omega}\right.\nonumber\\ 
~&& \left.~~-~ \frac{m^2a^2}{2\omega} \left(\mathcal{I}_{2}+\rs\mathcal{I}_{3}^{3}+\eta\mathcal{I}^{0}_{5}\right) \right\}~,
\end{eqnarray}
where, the expression of $\mathscr{I}^{\omega}_{lm}$ is given in \ref{eq:bound-GBF-m0}. The explicit evaluation of this quantity $\mathscr{I}^{\omega}_{lm}$ and the other quantities $\mathcal{I}_{2}$, $\mathcal{I}_{3}^{3}$, and $\mathcal{I}_{6}$ can be obtained from the evaluated integrals of \ref{Apn:Useful_rel}(in particular from \ref{eq:Def-integral-GBF}). Now it can be noticed that one can get the bound on the greybody factor for $m=0$ case(the expression obtained by putting the result of \ref{eq:bound-GBF-m0-f} in \ref{eq:GBF-bound-m0-1}) from these bounds of \ref{eq:bound-GBF-mn0-2} and \ref{eq:bound-GBF-mn0-3} by simply making $m=0$ in these expressions. Here the contribution of non zero $m$ comes through specific two quantities in the bound, and they also contain the effect of the deformation parameter $\eta$ for $m\neq0$. In \ref{fig:BGBF-etaV-DK-mLms} we have plotted this bound with respect to varying $\eta$ for fixed parameters $l=2$, $m=-1$, $a=0.05$ and different values of the frequency $\omega$. Note here we have kept the quantity $a\omega\ll1$ so that the approximation of \ref{eq:Spheriodal-Harmonics-eig} can be made. From this figure also one can observe that the bound on the greybody factor increases with an increasing value of the deformation parameter and mode frequency.

\subsubsection{super-radiant modes $m\ge\ms$}

The super-radiant modes are those for which $m\Omega_{H}>\omega$ or $m\ge\ms$. In this scenario the expression of the bound on the greybody factor from \ref{eq:bound-GBF-mn0-1} is further simplified using the triangle inequality and is expressed as
\begin{equation}\label{eq:theta-GBF-2}
 \scalebox{0.99}{$\Gamma(\omega) \geq 
\sech^2 \left\{\int_{-\infty}^{\infty}\left[\frac{|h'|}{2h}+ \frac{\mathbb{U}(r)}{2h} + \frac{|(h^2-\omega-m\Omega_{m})^2)|}{2h} \right]d\rstar\right\}~$},
\end{equation}\vspace{0.1cm}
where, $\mathbb{U}(r)$ is obtained from \ref{eq:potential-scalar-DK2}.
One can carry out this integral by considering two regions $m\in[\ms,2\ms)$ and $m\in[2\ms,\infty)$, where the value of $h$ can be taken to be positive and with a reasonable boundary condition, see \cite{Boonserm:2014rma}. Let us discuss these two situations in a case by case manner.
\vspace{0.15cm}

\emph{Case I $\left(m\in[\ms,2\ms)\right)$}:
In this case we consider the the function to be $h(r)=max\left\{\omega-m\Omega_{m},m\Omega_{H}-\omega\right\}$, which satisfies the requirement for $h(r)$ to be positive and also its asymptotic behaviors, see \cite{Boonserm:2014rma}. Then the first quantity of the right hand side of \ref{eq:theta-GBF-2} can be evaluated to be
\begin{equation}\label{eq:I1-mGms-1}
\int_{-\infty}^{\infty}\frac{|h'|}{2h} d\rstar = \left|\ln{h(r)}\right|_{\rh}^{\infty}=\ln{\left(\frac{\omega}{m\Omega_{H}-\omega}\right)}~.
\end{equation}
The second quantity in the same equation becomes 
\begin{equation}\label{eq:I2-mGms-1}
\int_{-\infty}^{\infty}\frac{\mathbb{U}(r)}{2h} d\rstar \le \int_{-\infty}^{\infty}\frac{\mathcal{U}(r)}{2(m\Omega_{H}-\omega)} dr=\frac{\mathscr{I}_{lm}}{2(m\Omega_{H}-\omega)}~,
\end{equation}
and the third quantity 
\begin{eqnarray}\label{eq:I3-mGms-1}
~&&\int_{-\infty}^{\infty}\frac{|h^2-(\omega-m\Omega_{m})^2)|}{2h} d\rstar = \mathcal{J}^{1}_{m}~,~(say)\nonumber\\
&=& \int_{-\infty}^{r_{0}}\frac{|(m\Omega_{H}-\omega)^2-(\omega-m\Omega_{m})^2)|}{2(m\Omega_{H}-\omega)} d\rstar~,
\end{eqnarray}
where, $r_{0}$ is obtained from equation $\omega-m\Omega_{m}(r_{0}) = m\Omega_{H}-\omega$.

\emph{Case II $\left(m\in[2\ms,\infty)\right)$}:
In this case the function can be chosen to be $h(r)=max\left\{m\Omega_{m}-\omega,\omega\right\}$, which satisfies the requirement for $h(r)$ to be positive and also its asymptotic behaviors, see \cite{Boonserm:2014rma}. Then the first quantity of the right hand side of \ref{eq:theta-GBF-2} can be evaluated to be
\begin{equation}\label{eq:I1-mGms-2}
\int_{-\infty}^{\infty}\frac{|h'|}{2h} d\rstar = \left|\ln{h(r)}\right|_{\rh}^{\infty}=\ln{\left(\frac{m\Omega_{H}-\omega}{\omega}\right)}~.
\end{equation}
The second quantity in the same equation becomes 
\begin{equation}\label{eq:I2-mGms-2}
\int_{-\infty}^{\infty}\frac{\mathbb{U}(r)}{2h} d\rstar \le \int_{-\infty}^{\infty}\frac{\mathcal{U}(r)}{2\omega} dr=\frac{\mathscr{I}_{lm}}{2\omega}~,
\end{equation}
and the third quantity 
\begin{eqnarray}\label{eq:I3-mGms-2}
~&&\int_{-\infty}^{\infty}\frac{|h^2-(\omega-m\Omega_{m})^2)|}{2h} d\rstar = \mathcal{J}^{2}_{m}~,~(say)\nonumber\\
&=& \int_{r_{0}^{'}}^{\infty}\frac{|\omega^2-(\omega-m\Omega_{m})^2)|}{2(m\Omega_{H}-\omega)} d\rstar~,
\end{eqnarray}
where, $r_{0}^{'}$ is obtained from equation $m\Omega_{m}(r_{0}^{'})-\omega=\omega$.

In these two cases also major contributions from \ref{eq:I2-mGms-1} and \ref{eq:I2-mGms-2}, in the bound on the greybody factor, are analogous to the results obtained from non super-radiant and $m=0$ cases. However, here the effects of deformation also come from the first and third quantities of the integral in \ref{eq:theta-GBF-2}. One may choose a suitable range of parameter values and depict this case also in a figure by plotting the lower bound on $\Gamma(\omega)$ with respect to $\eta$. However, that plot does not impart any new information.

\subsection{The bound on the greybody factor when $a=0$}\label{sub-sec:Schwarzschild-GBF-bounds}
The case of $a=0$ is particularly significant as it denotes the radially deformed static Schwarzschild BH spacetime. In this case one does not need to consider the intricacies coming from the super-radiant frequency, and the contribution from different angular momentum quantum number $m$ becomes irrelevant in the calculation of the bound on the greybody factor. Then one can consider the expressions provided for $m=0$ in \ref{eq:GBF-bound-m0-1} with \ref{eq:bound-GBF-m0} for the evaluation of these bounds.

\begin{figure}[htp]
\includegraphics[height=5cm,width=0.95\linewidth]{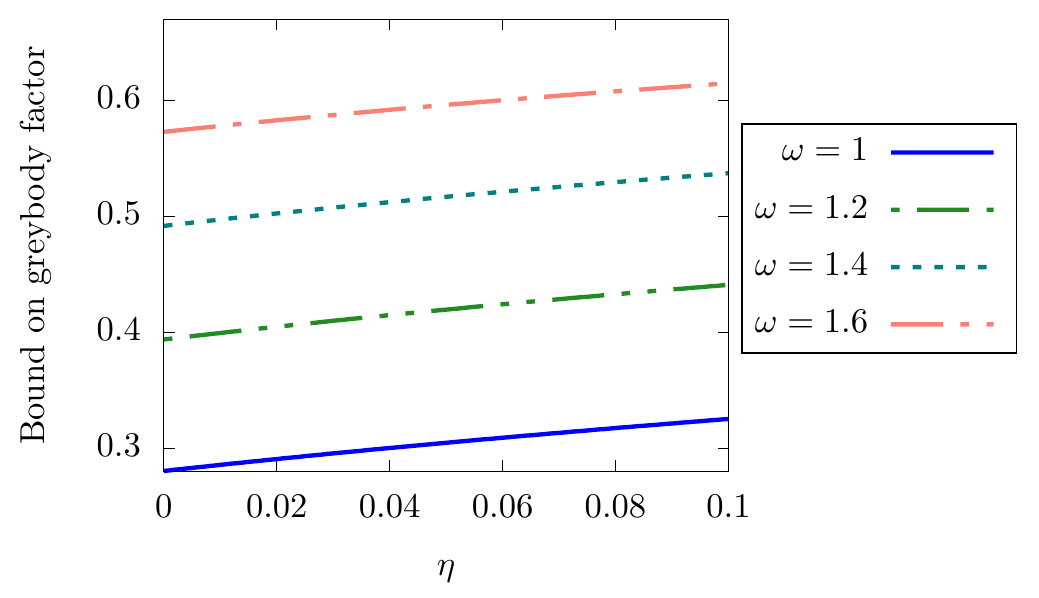}
\caption{In the above figure the lower bound on the greybody factor in a radially deformed Schwarzschild BH spacetime is plotted with respect to varying $\eta$ for different fixed $\omega$. We have set the other parameters at $l=1$, and $\rs=1$.}
\label{fig:BGBF-etaV-Sch}
\end{figure}

In particular, the expression of \ref{eq:bound-GBF-m0} in $a=0$ case can be evaluated to provide
\begin{eqnarray}\label{eq:bound-GBF-Schw1}
    \mathscr{I}^{\omega}_{lm} =\mathscr{I}_{lm} &=& \int_{\rh}^{\infty} dr \left[ \frac{l(l+1)}{r^2} 
+ \frac{\rs}{r^3} + \frac{3\eta}{r^5} \right]\nonumber\\
~&=& \frac{l(l+1)}{\rh} 
+ \frac{\rs}{2\rh^2} + \frac{3\eta}{4\rh^4}~,
\end{eqnarray}
where, we have used the fact that in the $a=0$ case $\mathcal{A}_{lm}^{\omega}$ denotes the eigenvalue corresponding to the Spherical harmonics equation, i.e., $\mathcal{A}_{lm}^{\omega}|_{a=0}=l(l+1)$, which can also be seen from \ref{eq:Spheriodal-Harmonics-eig}. Then the bound on the greybody factor can be estimated using \ref{eq:GBF-bound-m0-1} as
\begin{eqnarray}\label{eq:bound-GBF-Schw2}
 \Gamma(\omega) &\geq& 
\sech^2 \left\{\frac{1}{2\omega\rh} \left(l(l+1) 
+ \frac{\rs}{2\rh} + \frac{3\eta}{4\rh^3}\right) \right\}.
\end{eqnarray}

This ensures that at very high frequency the bound has the form
\begin{equation}
    \Gamma(\omega) \geq 
1- \frac{\left(l(l+1) 
+ \frac{\rs}{2\rh} + \frac{3\eta}{4\rh^3}\right)^2}{(2\omega\rh)^2} +\mathcal{O}\left[\frac{1}{\omega^4}
\right]~,
\end{equation}
in agreement with the result from Born approximation \cite{Visser:1998ke}. In \ref{fig:BGBF-etaV-Sch} we have plotted the bound of \ref{eq:bound-GBF-Schw2} and observed that in this radially deformed Schwarzschild BH spacetime also the value of the bound increases with increasing value of the deformation parameter and the frequency of the wave mode.

\section{Conclusion}\label{sec:conclusion}
In this article, we have discussed the thermal behavior of a BH spacetime which is deformed from the Kerr solution. We have mentioned that this deformation is only radial and therefore, the separability of scalar field wave equation and asymptotic properties remain the same as the Kerr spacetime. However, the position of the horizons and near horizon geometry differs from the Kerr case. This, in fact, motivated us to pursue the analysis concerning the Hawking radiation and bounds on the greybody factors in this spacetime.

The addition of the deformation parameter $\eta$ makes the horizon equation a cubic one, and there can be either one or three real positive solution(s), as seen in \ref{sec:horizon-structure-DK}. In this deformed BH spacetime, we observed that other than the event horizon, i.e., the outer horizon provided by the largest positive real root, there are other two inner horizons. This would make sure that the singularity is never naked, as it is covered by at least one horizon all the time. This is in stark contrast with Kerr BH case, where the extremality condition exists, and no real positive root of the horizon equation can be found for $a>\rs/2$.

The main objective of this article was to highlight how a radial deformation 
$\eta$ could change the thermal behavior of a BH solution.
In this regard, from \ref{Figure_03}, we observe that the 
height of the effective scalar field potential decreases with increasing $\eta$. 
It indicates more transmission of the field modes through the effective 
potential with increasing $\eta$, thus leading to the anticipation of an 
enhancement in the greybody factor.
In \ref{fig:SG-etaV} and \ref{fig:SG-aV-etaD}, the event horizon's surface 
gravity of this deformed BH spacetime is depicted for varying $\eta$, which 
imply an increasing Hawking temperature with increasing deformation parameter. 
In order to describe the bounds on the greybody factors in presence of $\eta$, 
which we discussed in \ref{sec:GBfactor}, we closely followed the analytical 
treatment as given in \cite{Boonserm:2008zg, Ngampitipan:2013sf, 
Boonserm:2014rma}. Note that these bounds are the lower bounds, while the upper 
bound is always $1$. In the case where bound becomes $1$, all the modes can 
escape to infinity unaltered, while the case of the lower bound nearing zero 
signify that much smaller amount of the wave modes transmit to the infinity. 
In \ref{sec:m_0_case} of \ref{sec:GBfactor}, we considered 
the $m=0$ case, and the bounds shown in \ref{fig:BGBF-etaV} and 
\ref{fig:BGBF-wV-etaD} imply an increase in the greybody factor for increasing 
deformation parameter $\eta$, which is consistent with the results obtained from 
\ref{Figure_03}. From these figures one also perceives that the greybody factor 
increases with increasing wave mode frequency $\omega$.
We observed in $m \neq 0$ case, a significant contribution in the bound coming 
from an integral \ref{eq:bound-GBF-m0} similar to the $m=0$ case, which keeps 
the dependence of the bounds on $\eta$ also same here, see 
\ref{fig:BGBF-etaV-DK-mLms}. Therefore, it is apparent that the radial 
deformation enhances the transmission probability for the modes to travel to 
infinity, which is also evident from \ref{Figure_03}. From this figure, we 
observed the height of the effective potential $\mathbb{V}(r)$ decreases with 
increasing the deformation parameter $\eta$, signifying a greater possibility 
for the field modes to transmit through the potential barrier. Moreover, from 
\ref{eq:Hawking-NumberDensity}, it can be observed that, at least for the $m=0$ 
case when the horizon's angular velocity does not contribute, the number density 
of Hawking quanta corresponding to a certain wave mode of frequency $\omega$ in 
a deformed BH spacetime will be higher compared to the Kerr case. Finally, in 
the last part of \ref{sec:GBfactor}, we found that for the Schwarzschild case 
also the bound on the greybody factor increases with increasing $\eta$. 

In passing, we would like to remind that a more general deformation can be found if we use $\rs \rightarrow \rs+\eta/r^{n+1}$ with $n \geq 1$, which also provides a spacetime where the field equation of motion is separable with respect to the radial and angular coordinates. Note that based on the value of $n$, different horizon structure emerges, and we end up with different number of horizons. However, as far as the thermal behavior is concerned, we are doubtful how much impact does $n$ can cause on the overall numbers. This is because as $n$ increases, the spacetime becomes more Kerr-like, and $\eta$ loses its essence. Therefore, the dominant contribution from $\eta$ would only come in the $n=1$ case, while for all $n >1$, the contribution from the deformation becomes dimmer. However, it remains an interesting arena to venture further.
\section*{Acknowledgement}
The authors acknowledge Bibhas Ranjan Majhi and Golam Mortuza Hossain for useful discussions concerning the current topic. S.M. wishes to thank Department of Science and Technology (DST),  Government of India, and S.B. thanks Indian Institute of Technology Guwahati (IIT Guwahati) for financial support.
\appendix
\labelformat{section}{Appendix #1} 
\labelformat{subsection}{Appendix #1}
\numberwithin{equation}{section}
\section{Surface gravity at the outer horizon}
\label{Apn:SurfaceGrav-OutHorizon}
In this part of the appendix we estimate the surface gravity at the outer horizon of the deformed BH. One can construct a vector null at the outer horizon \cite{book:Poisson} as
\begin{equation}\label{Eq:Apn-xi0}
    \xi^{\mu} = t^{\mu} + \Omega_{H}\phi^{\mu}~,
\end{equation}
where, $\Omega_{H}$ is the angular velocity of the outer horizon. The surface gravity $\kappa_{H}$ at the horizon can be obtained from the expression
\begin{equation}\label{Eq:Apn-SurfGrav0}
    \kappa~\xi_{\mu} = \frac{1}{2}\left(-\xi_{\nu}\xi^{\nu}\right)_{;\mu}~.
\end{equation}
In deformed spacetime this quantity $\xi_{\nu}\xi^{\nu}$ can be evaluated to be represented in terms of a simplified form given by
\begin{equation}\label{Eq:Apn-xixi}
    \xi_{\nu}\xi^{\nu} = \frac{\bar{\Sigma}\sin^{2}{\theta}}{\rho^2}\left(\Omega_{H}-\Omega\right)^2-\frac{\rho^2\bar{\Delta}}{\bar{\Sigma}}~,
\end{equation}%
where, the different quantities are given by 
\begin{eqnarray}\label{Eq:Apn-DK-BH-quantities}
    \bar{\Sigma} &=& (r^2+a^2)^2-a^2\bar{\Delta}\sin^2{\theta}~,\nonumber\\
   \textup{and,}~ \bar{\Delta} &=& (r^2+a^2-r\rs)-\frac{\eta}{r}~,
\end{eqnarray}%
and the expression of $\Omega$, which is the angular velocity of ZAMO, is taken from \ref{eq:AngularVel-ZAMO}. Now in \ref{Eq:Apn-xixi} the quantities $(\Omega_{H}-\Omega)$ and $\bar{\Delta}$ both vanishes at the outer horizon. In particular, one can express the quantity $\bar{\Delta}=(r-\rh) (r-r_{1}) (r-r_{2})$. Then the contribution of it in the surface gravity will be
\begin{equation}\label{Eq:Apn-xixi2}
    \scalebox{0.9}{$\left(-\xi_{\nu}\xi^{\nu}\right)_{;\mu} = \left(\frac{\rho^2\bar{\Delta}_{,r}}{\bar{\Sigma}} \right)\Big{|}_{r=\rh}\partial_{\mu}r = \left(\frac{\rho^2(r-r_{1}) (r-r_{2})}{\bar{\Sigma}}  \right)\Big{|}_{r=\rh}\partial_{\mu}r$}~.
\end{equation}
Furthermore, the vector $\xi_{\mu}$ can also be cast into the form $\xi_{\mu} = \left[\rho^2/(r^2+a^2)\right]|_{r=\rh}\partial_{\mu}r$ at the outer horizon. Then the surface gravity at the outer horizon can be readily found from the \ref{Eq:Apn-SurfGrav0} with the help of \ref{Eq:Apn-xixi2} as
\begin{equation}\label{Eq:Apn-SurfGrav1}
    \kappa = \frac{(\rh-r_{1})(\rh-r_{2})}{2\rh(\rh^2+a^2)}~.
\end{equation}
\vspace{0.2cm}

\section{Evaluation of the integrals necessary for estimating the bounds of \ref{sec:GBfactor}}
\label{Apn:Useful_rel}
In this part of the Appendix we evaluate the necessary integrals, which are used in \ref{eq:bound-GBF-m0-f} for the estimation of the greybody factor, and provide their explicit expressions.
\begin{eqnarray}
 \mathcal{I}_{1} &=& \int_{\rh}^{\infty} \frac{dr}{r^2+a^2} = \frac{\pi-2\tan^{-1}\left(\frac{\rh}{a}\right)}{2a}\nonumber
 \end{eqnarray}
 \begin{eqnarray}
  \mathcal{I}_{2} &=& \int_{\rh}^{\infty} \frac{dr}{(r^2+a^2)^2} = \frac{\pi-2\tan^{-1}\left(\frac{\rh}{a}\right)}{4a^3}-\frac{2a\rh}{4a^3 (\rh^2+a^2)}\nonumber
 \end{eqnarray}
 \begin{eqnarray}
  \mathcal{I}_{3}^{p} &=& \int_{\rh}^{\infty} \frac{r~dr}{(r^2+a^2)^p} = \frac{\left(\rh^2+a^2\right)^{1-p}}{2(p-1)}\nonumber
\end{eqnarray}
\begin{eqnarray}
\mathcal{I}^{p}_{4} &=& \int_{\rh}^{\infty} \frac{r~dr}{(r^2+a^2)^3r^p} = \frac{\rh^{-p-4}}{16} \left[-\frac{2pa^2 \rh^2+(p-2)\rh^{4}}{(\rh^2+a^2)^2}\right.\nonumber\\
~&& ~~\left. +~ \frac{2p(p+2)}{p+4}~ _{2}\mathcal{F}_{1}\left(1,\frac{p+4}{2},\frac{p+6}{2},-\frac{a^2}{\rh^2}\right)\right] \nonumber
\end{eqnarray}
\begin{eqnarray}\label{eq:Def-integral-GBF}
\mathcal{I}^{p}_{5} &=& \int_{\rh}^{\infty} \tfrac{dr}{(r^2+a^2)^3 r^{p+1}} = \tfrac{\rh^{-p-6}}{16}  \left[-\tfrac{2\rh^2(a^2(p+2)+p\rh^2)}{(\rh^2+a^2)^2}\right.\nonumber\\
~&+& \left.  \frac{2(p+2)(p+4)}{p+6}~ _{2}\mathcal{F}_{1}\left(1,\frac{p+6}{2},\frac{p+8}{2},-\frac{a^2}{\rh^2}\right)\right],\nonumber\\
\end{eqnarray}
where, $~_q\mathcal{F}_{s}(m;n;z)$ denotes the generalized \emph{Hyper geometric function}. It should also be noted that for our calculations the required expressions of the integrals $\mathcal{I}^{p}_{4}$, and $\mathcal{I}^{p}_{5}$ always have $p=0$.
\bibliographystyle{utphys1.bst}
\bibliography{bibtexfile}

\end{document}